\documentclass[prd,twocolumn,showpacs,groupedaddress,preprintnumbers,floatfix,letterpaper,superscriptaddress,nofootinbib]{revtex4}
\usepackage{graphicx,amssymb,amsmath,times,wrapfig,units,fancyhdr}
\usepackage[margin=10pt,font=small,labelfont=bf,flushleft]{caption}
\begin{document}

\preprint{FERMILAB-PUB-09-427-E, BNL-XXXXX-2009-XX}

\title{Observation of muon intensity variations by season with the MINOS far detector}


\newcommand{\Cambridge}{Cavendish Laboratory, University of Cambridge, Madingley Road, Cambridge CB3 0HE, United Kingdom}
\newcommand{\FNAL}{Fermi National Accelerator Laboratory, Batavia, Illinois 60510, USA}
\newcommand{\RAL}{Rutherford Appleton Laboratory, Science and Technology Facilities Council, Harwell Science and Innovation Campus, Didcot, OX11 0QX, United Kingdom}
\newcommand{\UCL}{Department of Physics and Astronomy, University College London, Gower Street, London WC1E 6BT, United Kingdom}
\newcommand{\Caltech}{Lauritsen Laboratory, California Institute of Technology, Pasadena, California 91125, USA}
\newcommand{\ANL}{Argonne National Laboratory, Argonne, Illinois 60439, USA}
\newcommand{\Athens}{Department of Physics, University of Athens, GR-15771 Athens, Greece}
\newcommand{\NTUAthens}{Department of Physics, National Tech. University of Athens, GR-15780 Athens, Greece}
\newcommand{\Benedictine}{Physics Department, Benedictine University, Lisle, Illinois 60532, USA}
\newcommand{\BNL}{Brookhaven National Laboratory, Upton, New York 11973, USA}
\newcommand{\CdF}{APC -- Universit\'{e} Paris 7 Denis Diderot, 10, rue Alice Domon et L\'{e}onie Duquet, F-75205 Paris Cedex 13, France}
\newcommand{\Cleveland}{Cleveland Clinic, Cleveland, Ohio 44195, USA}
\newcommand{\Delhi}{Department of Physics and Astrophysics, University of Delhi, Delhi 110007, India}
\newcommand{\GEHealth}{GE Healthcare, Florence South Carolina 29501, USA}
\newcommand{\Harvard}{Department of Physics, Harvard University, Cambridge, Massachusetts 02138, USA}
\newcommand{\HolyCross}{Holy Cross College, Notre Dame, Indiana 46556, USA}
\newcommand{\IIT}{Physics Division, Illinois Institute of Technology, Chicago, Illinois 60616, USA}
\newcommand{\Indiana}{Indiana University, Bloomington, Indiana 47405, USA}
\newcommand{\ITEP}{High Energy Experimental Physics Department, ITEP, B. Cheremushkinskaya, 25, 117218 Moscow, Russia}
\newcommand{\JMU}{Physics Department, James Madison University, Harrisonburg, Virginia 22807, USA}
\newcommand{\LASL}{Nuclear Nonproliferation Division, Threat Reduction Directorate, Los Alamos National Laboratory, Los Alamos, New Mexico 87545, USA}
\newcommand{\Lebedev}{Nuclear Physics Department, Lebedev Physical Institute, Leninsky Prospect 53, 119991 Moscow, Russia}
\newcommand{\LLL}{Lawrence Livermore National Laboratory, Livermore, California 94550, USA}
\newcommand{\MIT}{Lincoln Laboratory, Massachusetts Institute of Technology, Lexington, Massachusetts 02420, USA}
\newcommand{\Minnesota}{University of Minnesota, Minneapolis, Minnesota 55455, USA}
\newcommand{\Crookston}{Math, Science and Technology Department, University of Minnesota -- Crookston, Crookston, Minnesota 56716, USA}
\newcommand{\Duluth}{Department of Physics, University of Minnesota -- Duluth, Duluth, Minnesota 55812, USA}
\newcommand{\Otterbein}{Otterbein College, Westerville, Ohio 43081, USA}
\newcommand{\Oxford}{Department of Physics, University of Oxford, Oxford OX1 3RH, United Kingdom}
\newcommand{\Ohio}{Center for Cosmology and Astro Particle Physics, Ohio
  State University, Columbus, Ohio 43210, USA}
\newcommand{\Penn}{Department of Physics and Astronomy, University of Pennsylvania, Philadelphia, PA 19104-6396}
\newcommand{\PennState}{Department of Physics, Penn State University, State College PA 16802 }
\newcommand{\Pittsburgh}{Department of Physics and Astronomy, University of Pittsburgh, Pittsburgh, Pennsylvania 15260, USA}
\newcommand{\IHEP}{Institute for High Energy Physics, Protvino, Moscow Region RU-140284, Russia}
\newcommand{\RoyalH}{Physics Department, Royal Holloway, University of London, Egham, Surrey, TW20 0EX, United Kingdom}
\newcommand{\Carolina}{Department of Physics and Astronomy, University of South Carolina, Columbia, South Carolina 29208, USA}
\newcommand{\SLAC}{Stanford Linear Accelerator Center, Stanford, California 94309, USA}
\newcommand{\Stanford}{Department of Physics, Stanford University, Stanford, California 94305, USA}
\newcommand{\StJohnFisher}{Physics Department, St. John Fisher College, Rochester, New York 14618 USA}
\newcommand{\Sussex}{Department of Physics and Astronomy, University of Sussex, Falmer, Brighton BN1 9QH, United Kingdom}
\newcommand{\TexasAM}{Physics Department, Texas A\&M University, College Station, Texas 77843, USA}
\newcommand{\Texas}{Department of Physics, University of Texas at Austin, 1 University Station C1600, Austin, Texas 78712, USA}
\newcommand{\TechX}{Tech-X Corporation, Boulder, Colorado 80303, USA}
\newcommand{\Tufts}{Physics Department, Tufts University, Medford, Massachusetts 02155, USA}
\newcommand{\UNICAMP}{Universidade Estadual de Campinas, IFGW-UNICAMP, CP 6165, 13083-970, Campinas, SP, Brazil}
\newcommand{\USP}{Instituto de F\'{i}sica, Universidade de S\~{a}o Paulo,  CP 66318, 05315-970, S\~{a}o Paulo, SP, Brazil}
\newcommand{\Warsaw}{Department of Physics, University of Warsaw, Ho\.{z}a 69, PL-00-681 Warsaw, Poland}
\newcommand{\Washington}{Physics Department, Western Washington University, Bellingham, Washington 98225, USA}
\newcommand{\WandM}{Department of Physics, College of William \& Mary, Williamsburg, Virginia 23187, USA}
\newcommand{\Wisconsin}{Physics Department, University of Wisconsin, Madison, Wisconsin 53706, USA}
\newcommand{\deceased}{Deceased.}

\affiliation{\ANL}
\affiliation{\Athens}
\affiliation{\Benedictine}
\affiliation{\BNL}
\affiliation{\Caltech}
\affiliation{\Cambridge}
\affiliation{\UNICAMP}
\affiliation{\FNAL}
\affiliation{\Harvard}
\affiliation{\HolyCross}
\affiliation{\IIT}
\affiliation{\Indiana}
\affiliation{\UCL}
\affiliation{\Minnesota}
\affiliation{\Duluth}
\affiliation{\Otterbein}
\affiliation{\Oxford}
\affiliation{\Pittsburgh}
\affiliation{\RAL}
\affiliation{\USP}
\affiliation{\Carolina}
\affiliation{\Stanford}
\affiliation{\Sussex}
\affiliation{\TexasAM}
\affiliation{\Texas}
\affiliation{\Tufts}
\affiliation{\Warsaw}
\affiliation{\WandM}

\author{}%
\affiliation{}
\altaffiliation{\deceased}

\author{P.~Adamson}
\affiliation{\FNAL}

\author{C.~Andreopoulos}
\affiliation{\RAL}

\author{K.~E.~Arms}
\affiliation{\Minnesota}

\author{R.~Armstrong}
\affiliation{\Indiana}

\author{D.~J.~Auty}
\affiliation{\Sussex}


\author{D.~S.~Ayres}
\affiliation{\ANL}

\author{C.~Backhouse}
\affiliation{\Oxford}




\author{J.~Barnett}
\affiliation{\Oxford}

\author{G.~Barr}
\affiliation{\Oxford}

% ...

\collaboration{The MINOS Collaboration}
\noaffiliation

\date{\today}

\begin{abstract}
The temperature of the upper atmosphere affects the height of primary
cosmic ray interactions and the production of high-energy cosmic ray
muons which can be detected deep underground.
The MINOS far detector at Soudan MN, USA, has collected over 67 million cosmic ray induced muons.
The underground muon rate measured over a period of five years exhibits
a 4\% peak-to-peak seasonal variation which is highly correlated with the
temperature in the upper atmosphere.
The coefficient, $\alpha_T$, relating changes in
the muon rate to changes in atmospheric temperature
was found to be:  
$\alpha_T = 0.873~\pm~0.009$~(stat.)~$\pm 0.010$~(syst.).  
Pions and kaons in the primary hadronic interactions of
cosmic rays in the atmosphere contribute differently
to $\alpha_T$ due to the different masses and
lifetimes. This allows the measured value of $\alpha_T$
to be interpreted as a measurement of the K/$\pi$ ratio for $E_{p}\gtrsim$\unit[7]{TeV}
of $0.12^{+0.07}_{-0.05}$, consistent with the expectation from
collider experiments.
\end{abstract}


\maketitle

\section{Introduction }\label{sec:intro}
	
When very high energy cosmic rays interact in the stratosphere, mesons
are produced in the primary hadronic interaction. These mesons either interact and produce lower energy
hadronic cascades, or decay into high energy muons which can be
observed deep underground.  While the temperature of the troposphere
varies considerably within the day, the
temperature of the stratosphere remains nearly constant, usually changing
on the 
timescale of seasons (with the exception of the occasional Sudden
Stratospheric Warming~\cite{Osprey:2009}).  An increase in
temperature of the stratosphere causes a
decrease in density.  This reduces the chance of meson
interaction, resulting in a larger fraction decaying to produce muons.  This results in a higher muon rate
observed deep
underground~\cite{Barrett:1952,Ambrosio:1997tc,Bouchta:1999kg}. 
The majority of muons detected in the MINOS far detector are produced in 
the decay of pions, although the decays of kaons must be 
considered for a more complete description of the flux~\cite{Adamson:2007ww}. 

MINOS is a
long baseline neutrino oscillation experiment~\cite{Adamson:2007gu,Adamson:2005qc}, with a
neutrino source
and near detector at Fermi National Accelerator Laboratory in Batavia,
IL, USA, and a far detector at the Soudan
Underground Mine State Park in northern Minnesota, USA.  This paper describes
cosmic ray data taken in the far detector, a 
scintillator and steel tracking calorimeter
located \unit[0.72]{km} underground (\unit[2080]{mwe}, meters water
equivalent)~\cite{MinosNIM}.  It has a 
\unit[5.4]{kton} mass and a $\unit[6.91 \times 10^6] {cm^2
sr}$~\cite{Rebel:2004mm} acceptance.  
Because of its
depth, MINOS detects cosmic-ray muons with
energy at the surface, \unit[E$_{\mu}>$0.73]{TeV}. These high energy muons
are mostly the result of the decays of the mesons produced in
the primary hadronic interaction. This, coupled with the large
acceptance, makes it possible to detect small seasonal
temperature fluctuations in the upper atmosphere.
The far detector is the deepest underground detector with a magnetic field,
allowing the 
separation of particles by charge.  

The MINOS data are correlated
with atmospheric temperature measurements
 at the Soudan site provided by the European Centre for Medium-Range
Weather Forecasts (ECMWF)~\cite{ECMWF}.  This
temperature data set has higher precision than any other used for the
seasonal variation analysis~\cite{Barrett:1952,Bouchta:1999kg,Ambrosio:2002db,Cini:1967,Humble:1979,Cutler:1981,Sherman:1954,Fenton:1961,Andreyev:1987}.
The 67.32 million muon events used in this analysis were collected
over five years, from August~1,~2003 to July~31,~2008, a period that includes five complete annual
cycles.
The seasonal variations in muon intensity were compared to a theoretical model 
which extends the pion-only model of~\cite{Ambrosio:1997tc} 
to include the
contribution from kaons. 

\section{The Experimental Effective Temperature Coefficient}\label{sec:expint}
\subsection{Experimental Intensity}
The underground muon intensity depends on the threshold energy $E_\mathrm{th}$ and the
cosine of the zenith angle $\theta$.  The change in underground muon
intensity variations as a function of temperature was derived following
the formalism of~\cite{Barrett:1952,Ambrosio:1997tc}.
The change in the surface muon intensity, $\Delta
I_{\mu}(E,\cos \theta)$ occurring at the MINOS far detector site can
be written 
as:
\begin{equation}
\Delta I_{\mu} = \int_0^{\infty} \mathrm{d}X W(X) \Delta T(X)
\end{equation}
where $\Delta T(X)$ is the change in atmospheric temperature at
atmospheric depth $X$,
and the weight $W(X)$ reflects the temperature dependence of the
production of mesons in the atmosphere and their decay into muons that
can be observed in the far detector.  A temperature coefficient
$\alpha(X)$ can be defined as: 
\begin{equation}
\alpha(X) = \frac{T(X)}{I_{\mu}^0}W(X),
\end{equation}
where $I^{0}_{\mu}$ is the muon intensity evaluated at a given
value of atmospheric temperature
$T_0$. The phenomenological relationship between the atmospheric temperature
fluctuations and muon intensity variations can now be written
as:
\begin{equation}\label{eq:alpha}
\frac{\Delta I_{\mu}}{I^0_{\mu}} = \int_0^{\infty} \mathrm{d}X \alpha(X)
\frac{\Delta T(X)}{T_0}. 
\end{equation}

The atmosphere consists of many levels that vary continuously in both
temperature and pressure. To simplify calculations, the atmosphere is
approximated by an isothermal body with an effective temperature,
$T_\mathrm{eff}$, obtained from a weighted average over the atmospheric depth:
\begin{equation}\label{eq:teff}
T_\mathrm{eff} = \frac{\int_0^{\infty} \mathrm{d}X T(X) W(X)}  
{\int_0^{\infty} \mathrm{d}X  W(X)  } .
\end{equation}
An ``effective temperature coefficient'', $\alpha_{T}$ can then be defined
\begin{equation}\label{eq:teff_coeff}
\alpha_T = \frac{T_\mathrm{eff}}{I_{\mu}^0}\int_0^{\infty} \mathrm{d}X W(X).
\end{equation}
With these definitions in place, the relationship between atmospheric temperature
fluctuations and muon intensity variations can now be written as:
\begin{equation}\label{eq:temp_coeff}
\frac{\Delta I_{\mu}}{ I^0_{\mu}} = \alpha_{T} \frac{\Delta
  T_\mathrm{eff}}{T_\mathrm{eff}} .
\end{equation}
The configuration and geometric acceptance of the far detector remain
constant over time. Therefore, the rate, $R_{\mu}$ of muons observed in the
detector is proportional to the incident muon intensity and varies with
the effective atmospheric temperature as follows:
\begin{equation}
\frac{\Delta R_{\mu}}{\left < R_{\mu} \right >}~=~\alpha_{T}
\frac{\Delta T_\mathrm{eff}}{\left < T_\mathrm{eff} \right >} .
\end{equation}
In practice, the observed muon rates and the temperature data are averaged over
the period of a day. The effective temperature is obtained from a weighted average
of temperature measurements obtained at a set of discrete pressure levels.

The weight $W(X)$ can be written as the sum $W^{\pi} + W^{K}$, representing the
contribution of pions and kaons to the overall variation in muon intensity. The weights
$W^{\pi,K}$ are given by~\cite{Grashorn:2008dis,Grashorn:2009sm}:
\begin{equation}\label{eq:wval}
W^{\pi,K}(X)\simeq	
 \frac{  (1-X/ \Lambda_{\pi,K} ')^2  e^{-X/\Lambda_{\pi,K}} A^1_{\pi,K}
}{\gamma +\left( \gamma + 1 \right) B^1_{\pi,K} K(X)\left(\left<E_\mathrm{th}\cos\theta\right> / 
\epsilon_{\pi,K}\right)^2},
\end{equation}
where
\begin{equation}
K(X) \equiv \frac{(1- X / \Lambda_{\pi,K} ')^2}{(1-e^{-X / \Lambda_{\pi,K} '})
\Lambda_{\pi,K} ' /X}.
\end{equation}

The parameters $A^{1}_{\pi,K}$ include the amount of inclusive meson
production in the forward fragmentation region, masses of mesons and
muons, and muon spectral index; the input values
are $A^{1}_{\pi}$~=~1 and $A^{1}_{K}$~=~$0.38\cdot
r_{K/\pi}$~\cite{Grashorn:2008dis,Grashorn:2009sm}, where  $r_{K/\pi}$
is the K/$\pi$
ratio. 
The parameters $B^{1}_{\pi,K}$ reflect the relative atmospheric attenuation
of mesons; 
The threshold energy, $E_\mathrm{th}$,
is the energy required for a muon to survive to a particular depth;
The attenuation lengths for the cosmic ray primaries, pions and kaons are
$\Lambda_{N}$, $\Lambda_{\pi}$ and $\Lambda_{K}$ respectively
with $1/\Lambda^{'}_{\pi,K} \equiv 1/\Lambda_{N} - 1/\Lambda_{\pi,K}$.
The muon spectral index is given by $\gamma$.
The meson critical energy, $\epsilon_{\pi, K}$,
is the meson energy for which decay and interaction have an equal probability.
Since the distribution has a long tail (Fig.~\ref{fig:Spectra}), the
value of $\left<E_\mathrm{th}
\cos \theta\right>$ used here is the median.
The values for these parameters can be found in Table~\ref{tab:values}.
\begin{table}[!h]
\caption {Input W(X) parameter values.}
\label{tab:values}
\begin{tabular}{l l} \hline\hline
\hspace*{15pt}\textbf{Parameter} & \textbf{Value} \hspace*{15pt} \\\hline 
\hspace*{15pt}$A^1_{\pi}$ & 1~\cite{Grashorn:2008dis,Grashorn:2009sm} \\
\hspace*{15pt}$A^1_{K}$ & $0.38\cdot r_{K/\pi}$~\cite{Grashorn:2008dis,Grashorn:2009sm}\\
\hspace*{15pt}$r_{K/\pi}$ & 0.149~\cite{Gaisser:1990vg}~$\pm$~0.06~\cite{Barr:2006it}\\
\hspace*{15pt}$B^1_{\pi} $ & 1.460$\pm$ 0.007~\cite{Grashorn:2008dis,Grashorn:2009sm}\\
\hspace*{15pt}$B^1_{K} $ & 1.740 $\pm$ 0.028~\cite{Grashorn:2008dis,Grashorn:2009sm} \\
\hspace*{15pt}$\Lambda_{N}$ & \unit[120]{g/cm$^{2}$}~\cite{Gaisser:1990vg} \\
\hspace*{15pt}$\Lambda_{\pi}$ & \unit[180]{g/cm$^{2}$}~\cite{Gaisser:1990vg} \\
\hspace*{15pt}$\Lambda_{K}$ & \unit[160]{g/cm$^{2}$}~\cite{Gaisser:1990vg}\\
\hspace*{15pt}$\left<E_\mathrm{th} \cos \theta\right>$\hspace*{15pt} &\unit[0.785$\pm$0.14]{TeV}\\
\hspace*{15pt}$\gamma$ &  1.7$\pm$0.1~\cite{Adamson:2007ww} \\
\hspace*{15pt}$\epsilon_{\pi}$ & \unit[0.114$\pm$0.003]{TeV}~\cite{Grashorn:2008dis,Grashorn:2009sm}\\
\hspace*{15pt}$\epsilon_{K}$ & \unit[0.851$\pm$0.014]{TeV}~\cite{Grashorn:2008dis,Grashorn:2009sm}\\
\hline
\hline
\end{tabular}
\end{table}

\begin{figure}[!h]
\begin{center}
\includegraphics[width=0.5\textwidth]{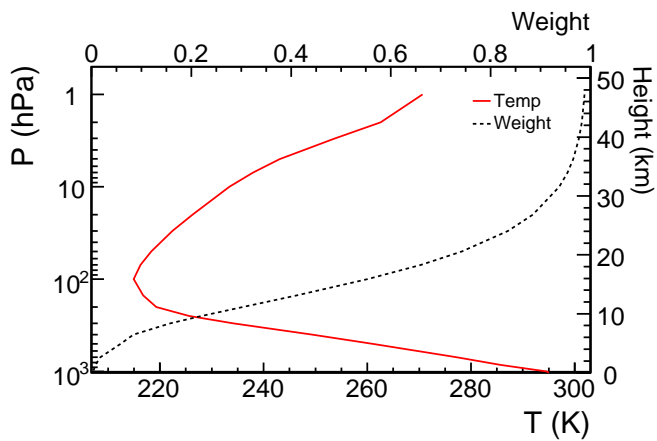}
\end{center}
\caption[Temperature Profile]{\label{fig:temp_profile}The five year average 
temperature at various pressure levels (solid line). The range is
from \unit[1000]{hPa} (\unit[1]{hPa}~=~\unit[1.019]{g/cm$^2$}), near
Earth's surface, to \unit[1]{hPa} (nearly \unit[50]{km}), near the top of 
the stratosphere.  The height axis on the right represents the average 
log-pressure height corresponding to the average temperatures plotted
here.  The dashed line is the weight as a 
function of pressure level ($X$) used to find $T_\mathrm{eff}$. 
The weights are determined by Eq.~\ref{eq:wval}, normalized to one. 
} 
\end{figure}

Since the temperature is measured at discrete levels, the integral
is represented by a sum
over the atmospheric levels $ X_n$: 
\begin{equation}\label{eq:teff_exp}
T_\mathrm{eff} \simeq \frac{\sum_{n=0}^{N} \Delta X_n
T(X_n)\left(W_n^{\pi}+W_n^{K}\right)}
{\sum_{n=0}^{N} \Delta X_n \left(W_n^{\pi}+W_n^{K}\right)} 
\end{equation}
where
$W^{\pi,K}_n $ is $W^{\pi,K} $  evaluated at $X_n$. 
The temperature and pressure vary
continuously through the atmosphere. Fig.~\ref{fig:temp_profile} (solid line)
shows the average temperature from 2003-2008 above Soudan as a function
of pressure level 
in the atmosphere~\cite{ECMWF}.
The height axis on the right represents the average  
log-pressure height, the height of a pressure
level relative to the surface pressure,  corresponding to the average temperatures plotted
here.  The dashed line is the weight as a 
function of pressure level $W(X)$, obtained from Eq.~\ref{eq:wval} and
normalized to one, used to calculate the effective temperature. 
The dashed weight curve in Fig.~\ref{fig:temp_profile} shows that the temperature fluctuations higher in the
atmosphere have a greater effect on the production of 
muons visible at a depth of \unit[2100]{mwe}. High energy mesons
produced at the top of the atmosphere are more likely to decay,
producing muons visible to MINOS, than those produced lower in the
atmosphere. Note that the expression used to calculate $T_\mathrm{eff}$ in the
pion scaling limit, ignoring the kaon contribution, is the same as the
MACRO calculation~\cite{Ambrosio:1997tc}.
The effective temperature coefficient, $\alpha_{T}$, is a function
of both the muon threshold energy and the K/$\pi$ ratio. As the energy
increases, the muon intensity becomes more dependent on the
meson critical energy, which in turn is proportional to the atmospheric
temperature. The effective temperature coefficient thus reflects the
fraction of mesons that are sensitive to atmospheric temperature
variations, and for energies much greater than the critical energy,
the value of $\alpha_{T}$ approaches unity. At the depth of the MINOS
far detector, the vertical muon threshold energy lies between the pion and kaon
critical energies. Therefore, because the muon energy is close to the
parent meson's energy, a larger K/$\pi$ ratio results in a
smaller value of $\alpha_{T}$.

\begin{figure}[!h]
\begin{center}
\includegraphics[width=0.5\textwidth]{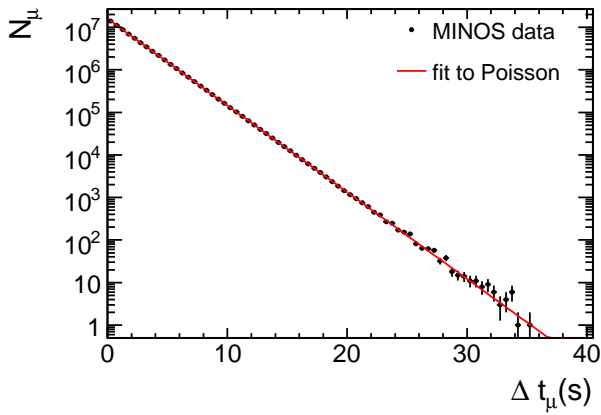}
\end{center}
\caption[Time Between Consecutive Muon Arrivals]{\label{fig:dt}The time between consecutive cosmic ray muon
arrivals, fit with a Poisson distribution.  The fit 
gives $\chi^2/N_{DoF}~=~55.2/68$; $\left<R_{\mu}\right>$~=~\unit[0.4692~$\pm$~0.0001]{Hz} (from slope).  The Poissonian  nature of the muon arrival times
 demonstrates the absence of short timescale systematic effects on the data.}
\end{figure}

\subsection{The Data}\label{sub:data}
The muon data for this analysis were accumulated over a five year span,
beginning on August 1, 2003.
Data quality cuts were performed to ensure a clean sample of muons
(Pre-Analysis cuts)~\cite{Adamson:2007ww} 
\begin{enumerate}
\item Require that all detector readout and sub-systems were functioning normally
\item Remove runs with anomalous cosmic ray rates, 
greater than \unit[1]{Hz}
\item Remove events that had many hits assigned to incorrect
  channels (properly demultiplexed~\cite{Adamson:2007ww})
\item Remove muons induced by NuMI beam interactions with timing cuts~\cite{Adamson:2007gu}.  
\end{enumerate}
After all cuts were 
applied the initial sample of 68.66 million muons was reduced to 67.32
million muons~\cite{Grashorn:2008dis}. 
A plot of the time between consecutive muon arrivals in the MINOS data is shown
in Fig.~\ref{fig:dt}.  The distribution is well described by a Poisson
distribution~\cite{Ahlen:1991yh,Grashorn:2008dis}
with mean rate $\left<R_{\mu}\right>$~=~\unit[0.4692~$\pm$~0.0001]{Hz},
demonstrating the absence of short-timescale systematic effects on the data. 
The average muon rate was calculated for each day by dividing the number of 
observed muons by the detector livetime.  

\begin{figure}[!h]
\begin{center}
\includegraphics[width=0.5\textwidth]{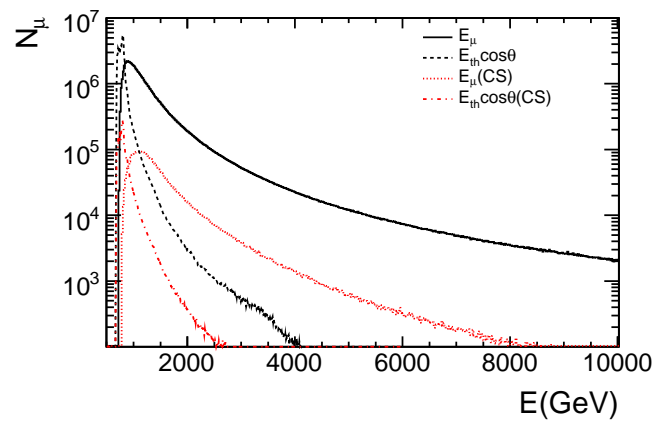}
\caption[Energy Spectra]{\label{fig:Spectra}
A plot of the energy spectra observed in the far detector.  The dashed
line is $E_\mathrm{th}\cos\theta$, which was used to determine the
value used in Eq.~\ref{eq:wval}.  The solid line is the distribution of
muon surface energies, $E_{\mu}$ in far detector.
 Also shown are 
$E_\mathrm{th}\cos\theta(CS)$ (dot-dash line), the distribution of
$E_\mathrm{th}\cos\theta$ after charge-separation cuts have been
applied, and
$E_{\mu}(CS)$ (dotted line), the distribution of $E_{\mu}$ after the
charge-separation cuts (see Sec.~\ref{sub:chargesep}) have been
applied.  Note that the charge-separation cuts have been applied, but
the distributions shown include both muon species.
}
\end{center}
\end{figure}

The energy spectra for the observed muons can be seen in
Fig.~\ref{fig:Spectra}.  
 The solid
line is $E_\mathrm{th}\cos\theta$, which was used to determine the
value used in Eq.~\ref{eq:wval}.  The dashed line is the distribution of
muon surface energies,
$E_{\mu}$, in far detector, which has a much longer
tail than the distribution of threshold energies.  Also shown are 
$E_\mathrm{th}\cos\theta(CS)$ (dot-dash line), the distribution of
$E_\mathrm{th}\cos\theta$ after charge-separation cuts have been
applied, and
$E_{\mu}(CS)$ (dotted line), the distribution of $E_{\mu}$ after the
charge-separation cuts (see Sec.~\ref{sub:chargesep}) have been
applied.  Note that the charge-separation cuts have been applied, but
the distributions shown include both muon species.  The  $E_\mathrm{th}\cos\theta$ distribution is
peaked, with a median value
$\left<E_\mathrm{th}\cos\theta\right>$\unit[=0.795$\pm$0.14]{TeV}.  This
distribution, with its rapid fall-off, reflects the rock overburden surrounding the far detector.

The temperature data for the Soudan site was obtained from 
ECMWF, which collates a number 
of different types of observations (e.g. surface, satellite and upper air sounding)
at many  locations around
the globe, and uses a global atmospheric model to interpolate to a particular
location. For this analysis, the ECMWF 
model produced atmospheric temperatures at 21 discrete pressure levels:
1000, 925, 850, 700, 500, 400, 300, 250, 200, 150, 100, 70, 50, 30, 20,
10, 7, 5, 3, 2
and \unit[1]{hPa} (\unit[1]{hPa}~=~\unit[1.019]{g/cm$^2$}), at four
times, \unit[0000]{h}, \unit[0600]{h}, \unit[1200]{h} and \unit[1800]{h} each day.  
The effective temperature, $T_\mathrm{eff}$, was
calculated four times 
each day using Eq.~\ref{eq:teff_exp}. A mean value
$\left<T_\mathrm{eff}\right>$ and error was obtained 
from these four daily measurements.
The ECMWF temperature data was cross-checked using the Integrated Global
Radiosonde Archive (IGRA) of temperature measurements~\cite{IGRA:2006}.
The distribution of the differences between ECMWF and IGRA temperature values at
International Falls, MN was well described by  a Gaussian distribution
with $\sigma$~=~\unit[0.31]{K}.

\begin{figure}[!h]
\begin{center}
%
\includegraphics[width=0.5\textwidth]{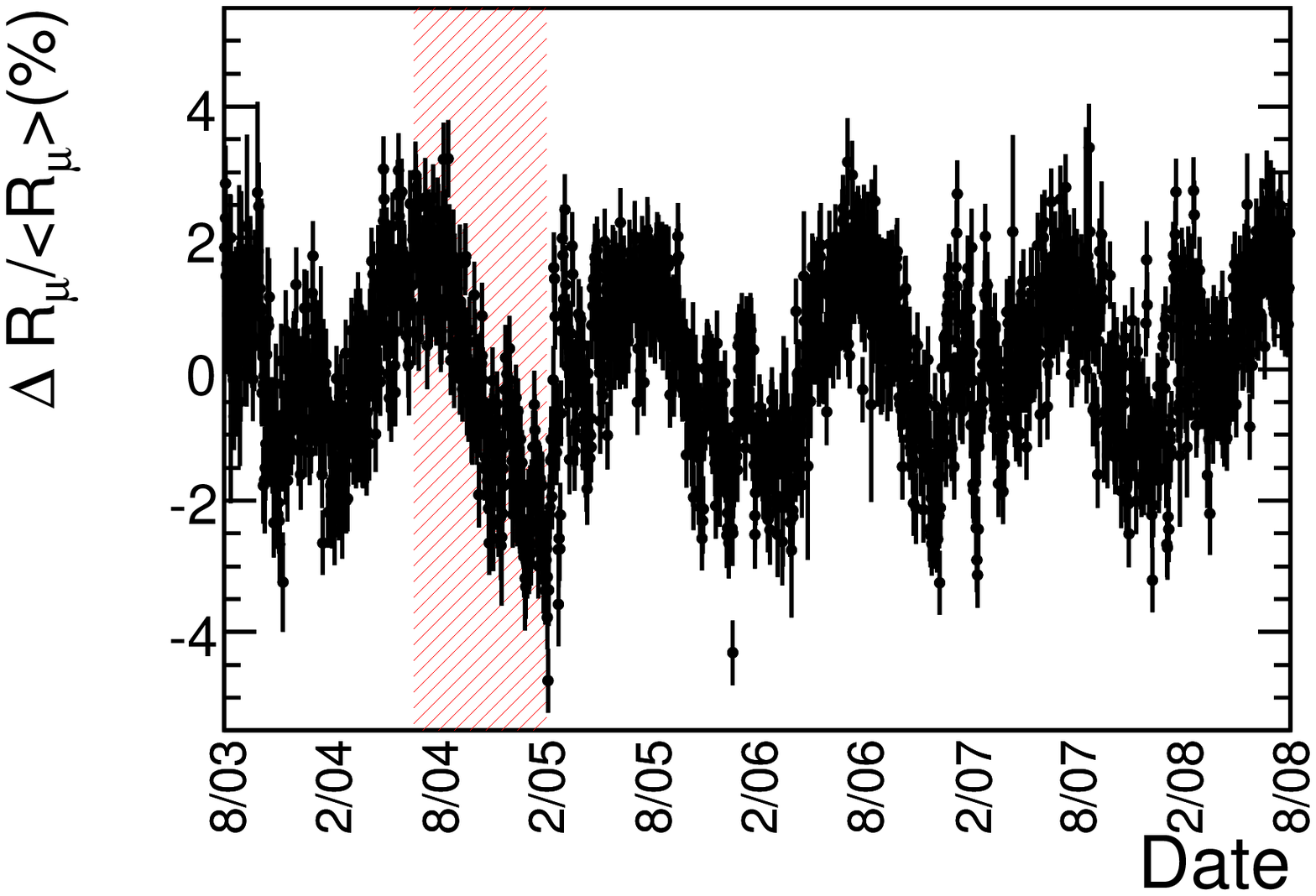}
\caption[Daily Muon Rate]{\label{fig:seasonal}The daily deviation from
  the mean rate of cosmic 
ray muon arrivals from 8/03-8/08, shown here with statistical error bars.  The periodic fluctuations have the expected
maxima in August, minima in February. The hatched region indicates the
period of time when the detector ran with the magnetic field reversed
from the normal configuration.}
\includegraphics[width=0.5\textwidth]{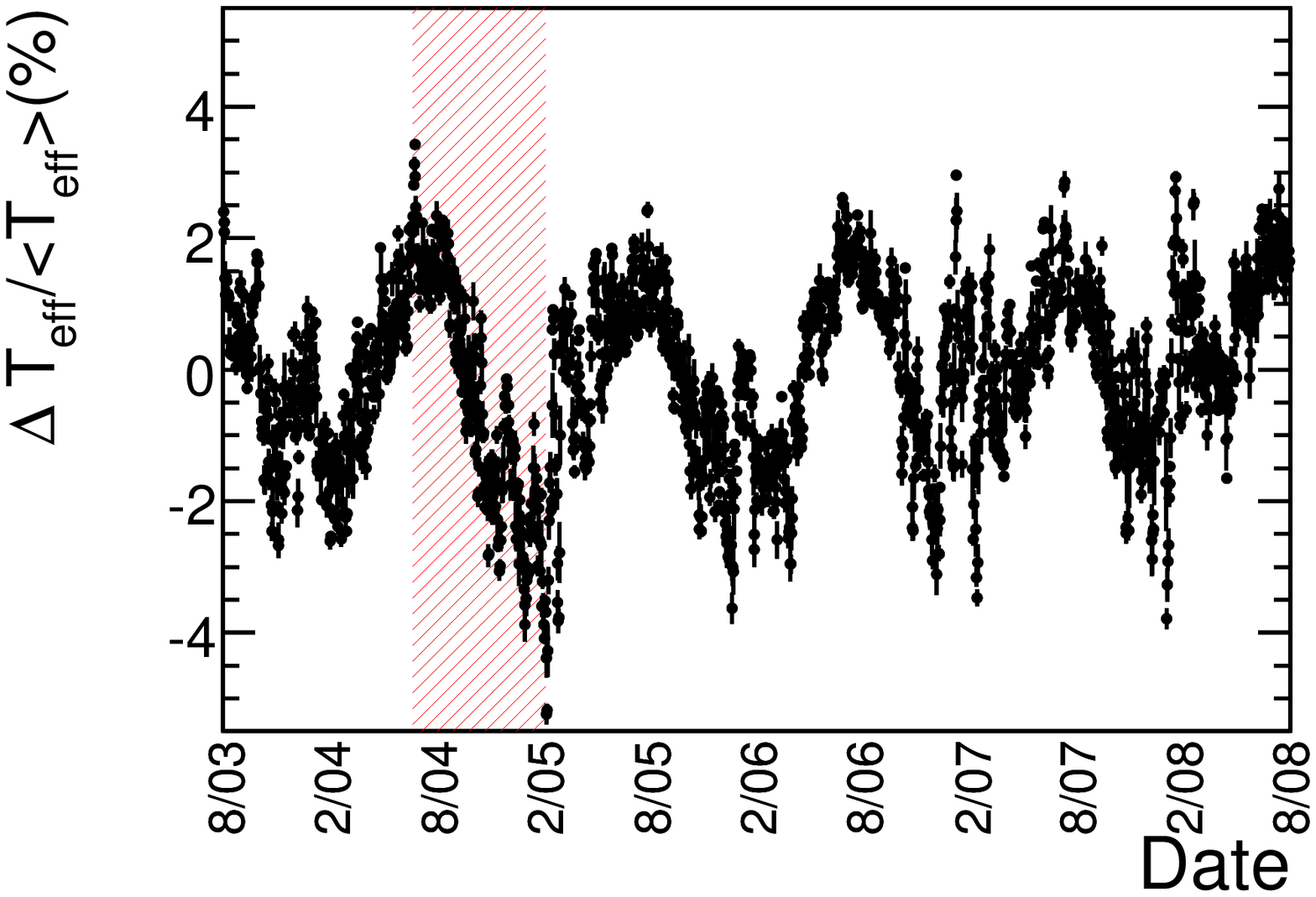}
\caption[Daily Effective Temperature]{\label{fig:teff}The daily
deviation from the mean effective temperature 
over a period of five years, beginning when the far
detector was complete, 08/03-08/08.  The hatched region indicates the
period of time when the detector ran with the magnetic field reversed
from the normal configuration.}
\end{center}
\end{figure}

Fig.~\ref{fig:seasonal} shows the percentage deviation in the mean daily muon rate,
$\Delta R_{\mu}$, over the entire set of data, with statistical error bars. 
A typical day at $\unit[\left<R_\mu \right>~=~0.4692]{Hz}$ yields $\sim$40,000 muons,
resulting in  error bars of order 0.5\%.
The variation with season can be seen, with maxima in August and minima
in February.  These maxima peak at rates that are within 0.5\% of each other. 
For the five year period $\left<T_\mathrm{eff}\right>~=~\unit[221.93]{K}$.  
The distribution of $\Delta T_\mathrm{eff}$ over the data period can be seen in
Fig.~\ref{fig:teff}, with strong periodic seasonal correlation with the data.   There
is also striking correspondence between Fig.~\ref{fig:seasonal} and
Fig.~\ref{fig:teff} for small term maxima and minima over a few days' span.

\begin{figure}[!h]
\begin{center}
\includegraphics[width=0.5\textwidth]{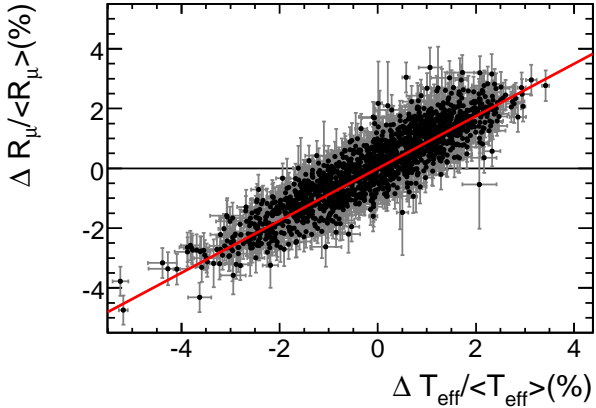}
%
\caption[Time Series R(T)]{\label{fig:RTCorrP}
A plot of 
$\Delta R_{\mu}/\left<R_{\mu}\right>$ as a function of $\Delta T_\mathrm{eff}/\left<T_\mathrm{eff}\right>$ for
single muons, fit by a line with the y-intercept fixed at 0.  The fit has a $\chi^2/N_{DoF} =
1959/1797$, and the slope is $ \alpha_T~=~0.873~\pm~0.009$.
}
\end{center}
\end{figure}

 A plot of
$\Delta R_{\mu}/\left<R_{\mu}\right>(\Delta T_\mathrm{eff})$ was produced (Fig.~\ref{fig:RTCorrP}) for
each day's $\Delta R_{\mu}$ and $\Delta T_\mathrm{eff}$ data to quantify
the daily correlation between rate and temperature.
To find the value for $\alpha_T$,  a linear
regression was performed using the MINUIT~\cite{James:1975dr} fitting
package. This package
performs a linear regression accounting for error bars on both the x and
y axis using a numerical minimization method.  The result of this fit is
a slope of $\alpha_T~=~0.873~\pm~0.009$ (statistical errors only), and the correlation coefficient
(R-value) between these two distributions is 0.90.  

The effects of systematic uncertainties were evaluated by modifying parameters and
recalculating $\alpha_T$. Table~\ref{tab:errors}  shows the difference in calculated
$\alpha_T$ for the modified
parameters. The largest systematic errors are: a) the $\pm$~0.06 uncertainty in meson
production ratio~\cite{Barr:2006it}; b) the \unit[$\pm0.31$]{K} uncertainty 
in mean effective temperature, estimated by
comparing ECMWF temperatures at International Falls, MN, to those of the IGRA~\cite{IGRA:2006}
measurements; c) the \unit[$\pm$0.14]{TeV} uncertainty in muon threshold
energy, estimated from uncertainties in the rock overburden above the
far detector.  To estimate this uncertainty, the rock map was adjusted
up by 10\% and $\left<E_\mathrm{th}\cos\theta \right>$ was
calculated, then the rock map was adjusted down by 10\% and
$\left<E_\mathrm{th}\cos\theta \right>$ was again recalculated.
The uncertainty was then calculated from the difference between
$\left<E_\mathrm{th}\cos\theta \right>$ and these adjusted
values.  
\begin{table}[!h]
\caption {Systematic errors on the experimental parameter inputs to $\alpha_T$ .}
\label{tab:errors}
\begin{tabular}{l l} \hline\hline
\hspace*{15pt}\textbf{Parameter} & $\Delta \alpha_T$ \hspace*{15pt} \\\hline 
\hspace*{15pt}meson production ratio, $r_{K/\pi}$ =
0.149$\pm$0.06~\cite{Barr:2006it}
\hspace*{15pt} & 0.007 \\  
\hspace*{15pt}mean effective temperature, $\left<T_\mathrm{eff} \right>$= 
\unit[221.93$\pm$0.32]{K} 
\hspace*{5pt} & 0.0051 \\ 
\hspace*{15pt}threshold energy, $\left<E_\mathrm{th} \cos \theta
 \right>$=\unit[0.795$\pm$0.14]{TeV} 
 & 0.0048\\
\hspace*{15pt}kaon constant, $B^1_{K} $~=~1.740 $\pm$ 0.028 
& 0.00046 \\
\hspace*{15pt}pion constant, $B^1_{\pi} $~=~1.460 $\pm$ 0.007  
& 0.000063 \\
\hline
\hspace*{0.75cm}\textbf{Total} 
 & \textbf{0.010}  
\\\hline\hline
\end{tabular}
\end{table}
These systematic errors were added in quadrature and are included with
the error from the linear fit to obtain the experimental value of
$\alpha_T~=~0.873~\pm~0.009$~(stat.)~$\pm~0.010$~(syst.).

\subsection{Charge Separated}\label{sub:chargesep}
To obtain a sample of events with well-measured charge sign,
further selection requirements were applied to the length and radius
of curvature of muon tracks. These cuts, taken from previous
investigations of the muon charge ratio at MINOS~\cite{Adamson:2007ww}, have
the effect of reducing the energy distribution at Earth's surface of the
selected muon sample.

In all, 5.7\% of the data set survived the cuts for both the forward and
reverse field detector configurations. 
For the charge-separated samples linear
regressions yielded effective temperature coefficients,
$\alpha_T(\mu^+)~=~0.79~\pm~0.05$ and $\alpha_T(\mu^-)~=~0.77~\pm~0.06$ with
$\chi^2$/$N_{DoF}$ of 1933/1758 and 1688/1751
respectively. These numbers are consistent with each other,
so there is no measurable difference between the
temperature effect on $\mu^+$ and $\mu^-$.
The value of the charge-separated $\alpha_T$ is expected to be smaller
than the previous $\alpha_T$ with no charge separation because the
selection cuts change the energy distribution over which the integration
is performed to calculate $\alpha_T$.  This can be seen in
Fig.~\ref{fig:Spectra}, with the most 
dramatic difference between the all muon and charge separated
distributions of $E_{\mu}cos\theta$.  
This difference could 
produce the systematic offset observed between these values, and is
discussed further in the next section.


\section{Discussion}

\subsection{Predicted $\alpha_T$}\label{sub:theoretical_alpha}
The theoretical prediction of $\alpha_T$  can be written as
~\cite{Barrett:1952}: 
\begin{equation}\label{eq:diffalpha}
\alpha_T
~=~- \frac{E_\mathrm{th}}{I^0_{\mu}}\frac{\partial I_{\mu}}{\partial E_\mathrm{th}} -
 \gamma
\end{equation}

Using the differential muon intensity~\cite{Gaisser:1990vg},
\begin{eqnarray}\label{eq:energyspectrum}
\frac{dI_{\mu}}{dE_{\mu}} &=& \int_0^{\infty}\mathcal{P}_{\mu}(E,X)dX
\simeq
A \times E^{-(\gamma+1)}
\nonumber
\\
&\times&\left(\frac{1}{1+1.1E_{\mu}\cos
\theta/\epsilon_{\pi}}+\frac{0.38r_{K/\pi}}{1+1.1E_{\mu}\cos
\theta/\epsilon_{K}}\right),
\end{eqnarray}
and the MACRO approximation for the muon
intensity~\cite{Ambrosio:1997tc}, the prediction for $\alpha_T$ can be calculated:
\begin{eqnarray}\label{eq:thalpha}
\alpha_T &=&
\frac{1}{D_{\pi}}
\frac{1/\epsilon_{K}+A^1_K(D_{\pi}/D_{K})^2 /\epsilon_{\pi} } {
  1/\epsilon_{K} + A^1_K(D_{\pi}/D_{K})/\epsilon_{\pi}
} 
\end{eqnarray}
where
\begin{equation}\label{eq:dpik}
D_{\pi,K}~=~\frac{\gamma}{\gamma+1} \frac{ \epsilon_{\pi,K}}{1.1 E_\mathrm{th} \cos
\theta } + 1,  
\end{equation}
Note that this can be reduced to MACRO's previously
published expression $\left( \alpha_T \right)_{\pi}$
\cite{Ambrosio:1997tc}, by
setting $A^1_K=0$ (no kaon contribution).  $A^1_{K}~=~0.38 \cdot
r_{K/\pi}$ is the same as in Sec.~\ref{sec:expint}.

A numerical integration using a Monte Carlo 
method was performed to find the predicted value of the seasonal effect
coefficient,
 $\left< \alpha_T \right>_\mathrm{p}$, for the far detector. A set of
muons was generated by drawing values of $E_{\mu}$
and $\cos\theta$ separately from the differential intensity of muons 
at the surface, calculated in~\cite{Gaisser:1990vg}. A random 
azimuthal angle, $\phi$, was assigned to each event and combined with  
$\cos\theta$ and the Soudan rock overburden map~\cite{Adamson:2007ww}
to find the slant depth, $S(\cos\theta, \phi)$, of the event. 
This was converted into the corresponding threshold energy, $E_\mathrm{th}$,
required for a muon on the surface to propagate to the far
detector. Events satisfying $E_{\mu} > E_\mathrm{th}$ were 
retained, and the mean value of $\alpha_T$ was found for a sample of
10,000 events, giving $\left<\alpha_T\right>_\mathrm{p}~=~0.864~\pm~0.024$ for MINOS. 
When this calculation was performed using the lower energy charge-separated energy 
spectrum, the result is an $\left<\alpha_T\right>_\mathrm{p}$ value that is
lower by 0.015.  This is most clearly seen in Eq.~\ref{eq:thalpha},
which is dominated by the leading $1/D_{\pi}$ term.  As
$E_\mathrm{th}cos\theta$ increases, $ D_{\pi}$ goes to one.  Any
selection that reduces the $E_\mathrm{th}cos\theta$ distribution will
then reduce the expected $\alpha_T$.

The systematic uncertainty on $\left<\alpha_T\right>_\mathrm{p}$ was found by
modifying the input parameters and recalculating $\alpha_T$. The dominant
contributions were from: a) the $\pm$~0.06 uncertainty in meson production
ratio; b) the $\pm$~10\% in rock map uncertainty\footnote[1]{ The rock
map is not a determination of the slant depth by geophysical means.
It was created by measuring the muon flux coming from a particular
solid angle region on the sky and then normalizing to the All-world
Crouch underground muon intensity curve~\cite{Crouch:1987}.  This was done with both
Soudan~2 data~\cite{Kasahara:1997} and with MINOS
data~\cite{Adamson:2007ww}, and these calculations were shown to agree
to within 10\%.  Average cosmic ray muon flux, like those determined here and
in~\cite{Adamson:2007ww}  can be determined using this method, although in any particular
direction the rock map can be much different from what was calculated
(e.g., in the direction of iron veins).}; c) the $\pm$~0.1 uncertainty in
muon spectral index;
d) the \unit[$\pm$~0.014]{TeV} uncertainty in kaon critical energy; and e)
  the \unit[$\pm$~0.003]{TeV} uncertainty in pion critical energy. 
These uncertainties are summarized in
Table~\ref{tab:therrors}. 
\begin{table}[!h]
\caption {Systematic errors on the theoretical parameter inputs to $\alpha_T$.}
\label{tab:therrors}
\centering
\begin{tabular}{l l} \hline\hline
\hspace*{15pt}\textbf{Parameter}  & 
$\Delta \alpha_T $ \hspace*{15pt} \\\hline
\hspace*{15pt}meson production ratio, K/$\pi$ =
0.149$\pm$0.06~\cite{Barr:2006it}
\hspace*{15pt} & 0.020 \\
\hspace*{15pt}rock map uncertainty  $\pm10\%$ 
 &0.013\\
\hspace*{15pt}muon spectral index, $\gamma$= 1.7 $\pm$ 0.1
&  0.0031 \\ 
\hspace*{15pt}kaon critical energy,
$\epsilon_{K}$=\unit[0.851$\pm$ 0.014]{TeV} \hspace*{20pt}  
& 0.0014\\
\hspace*{15pt}pion critical energy, $\epsilon_{\pi}$=\unit[0.114$\pm$0.003]{TeV} 
& 0.0002 \\
\hline
\hspace*{0.75cm}\textbf{Theoretical Total}  
&\textbf{0.024}    \\ \hline \hline  
\end{tabular}
\end{table}

\begin{figure}[h]
  \begin{center}
\includegraphics[width=0.5\textwidth]{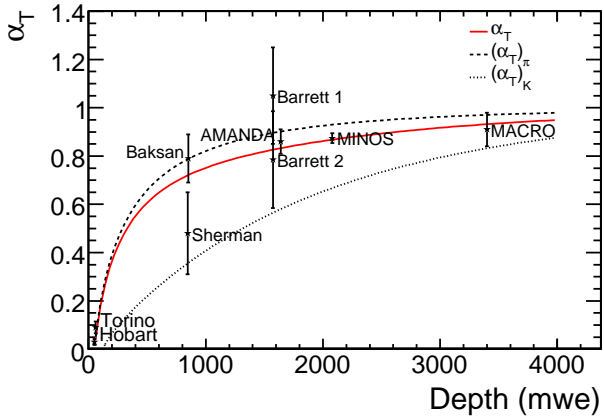}
\caption{\label{fig:alphaGlobal} 
 The theoretical prediction for $\alpha_T$ as a function of detector
 depth. The dashed (top) curve is the prediction using the pion-only
 model (of MACRO) and the dotted (bottom) curve is the prediction
 using a kaon-only model. The solid (middle) curve is the new
 prediction including both K and $\pi$.  These curves are illustrative
 only as the definition of effective temperature used to calculate the
 experimental values also depends on the K/$\pi$ ratio.
 The data from
 other experiments are shown for comparison only, and are from 
 Barrett~1,~2~\cite{Barrett:1952}, AMANDA~\cite{Bouchta:1999kg},
 MACRO~\cite{Ambrosio:2002db}, Torino~\cite{Cini:1967},
 Sherman~\cite{Sherman:1954}, Hobart~\cite{Fenton:1961}  and Baksan~\cite{Andreyev:1987}.} 
\end{center}
\end{figure}
Fig.~\ref{fig:alphaGlobal} shows effective temperature coefficients
from MINOS and other underground 
experiments, including those of the MACRO survey~\cite{Ambrosio:1997tc},
as a function of detector depth. 
The MINOS and Sherman~\cite{Sherman:1954} effective temperature coefficients
shown in Fig.~\ref{fig:alphaGlobal} were calculated using
Eq.~\ref{eq:teff_exp}.  The other experimental data points are taken from
the MACRO survey~\cite{Ambrosio:1997tc}
and were calculated using a definition which excluded the contributions from kaons and were limited by
temperature measurements up to \unit[20]{g/cm$^2$}; when the MINOS result is recalculated with this
definition the effective temperature coefficient decreases to $\alpha_T$ = 0.835.
To compare the experimental values with the theoretical model,
Eq.~\ref{eq:thalpha}, the expected 
effective temperature coefficient as a function of depth was calculated
using the numerical integration method 
outlined earlier, using standard rock and a flat 
overburden, and is shown in 
Fig.~\ref{fig:alphaGlobal} as the solid line. 
There is qualitative agreement between the prediction and the
experimentally measured 
values, but quantitative comparisons would require recalculating the
experimental values 
using the kaon-inclusive definition of effective temperature. The two
dashed lines in Fig.~\ref{fig:alphaGlobal} show the effective temperature dependence for the extreme pion-only
and kaon-only 
predictions.  Fig.~\ref{fig:alphaGlobal} is illustrative only, as the dependence of the
experimentally measured effective temperature coefficient on the input K/$\pi$ ratio is not explicitly
shown.
\subsection{Measurement of Atmospheric K/$\pi$ Ratio}\label{sub:kpi}
The 
uncertainty on the atmospheric K/$\pi$ ratio in the current cosmic ray flux 
models is of order
\unit[40]{\%}~\cite{Barr:2006it}.  There has not been a measurement of
this ratio with cosmic rays. 
Previous measurements have been made at
accelerators for p+p collisions~\cite{Rossi:1974if}, Au+Au
collisions~\cite{Adler:2002wn}, Pb+P
collisions~\cite{Afanasiev:2002mx,Alt:2005zq} and p+$\bar{\mathrm{p}}$ collisions~~\cite{Alexopoulos:1993wt}. Many other older
measurements are summarized in \cite{Gazdzicki:1995zs}.
The experimental and theoretical values of $\alpha_{T}$ can be combined
to give a new measurement of the K/$\pi$ ratio for the reaction
$p+A_\mathrm{atm}$, with $E_p\gtrsim$\unit[7]{TeV}.  The threshold muon surface
energy, \unit[$E_\mathrm{th}$=0.73]{TeV} and the median 
muon surface energy, $\left<E_{\mu}\right>$, is \unit[1.15]{TeV}.  On
average, the muon energy is one tenth the energy of its parent primary.  
The theoretical $\alpha_{T}$
depends directly on the K/$\pi$ ratio, as a consequence of the different
interaction and decay properties of kaons and pions in the atmosphere.
Since kaons and pions have different critical energies and attenuation
lengths, the effective temperature also depends on the K/$\pi$ ratio,
and therefore the experimental $\alpha_{T}$ is a weak function of the
K/$\pi$ ratio. By plotting the experimental and theoretical values of
$\alpha_{T}$ as functions of the K/$\pi$ ratio and finding the intersection
of the two curves, a measurement of the K/$\pi$ ratio can be obtained.

\begin{figure}[!h]
\begin{center}
\includegraphics[width=.5\textwidth]{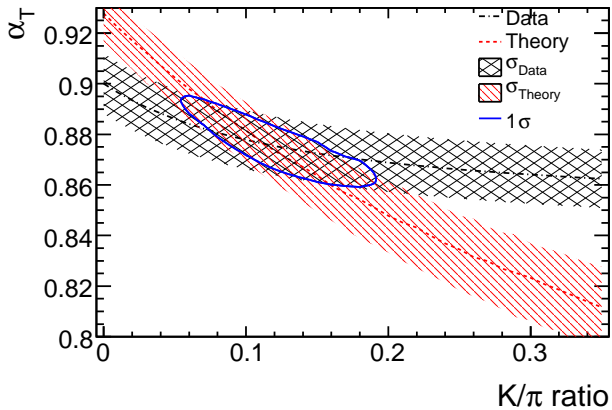}
\end{center}
\caption[ $\alpha_T$ as a Function of the K/$\pi$ ratio]{\label{fig:alphaTKPi}
The MINOS experimental $\alpha_T$ as a function of the K/$\pi$ ratio
(dot-dash
line), with its error given by the cross-hatched region, on the same axes as the theoretical $\alpha_T$ as a function
of the K/$\pi$ ratio (dashed line), with its error given by the hatched
region. 
The error on the experimental $\alpha_T$ (from
Table~\ref{tab:errors} $E_\mathrm{th} \cos \theta$,  $B^1_{\pi,K}$ and  
 $\left<T_\mathrm{eff} \right>$) plus statistical error is $\pm~0.012$, and the
theoretical 
$\alpha_T$ error (from $\epsilon_{\pi,K}$ and the rock map,
Table~\ref{tab:therrors}) is $\pm~0.013$ at the best fit point.  The
intersection is at K/$\pi$=$0.12^{+0.07}_{-0.05}$. The solid line denotes the
1$\sigma$ contour around the best fit. }
\end{figure}

Fig.~\ref{fig:alphaTKPi} shows the experimental and theoretical values of $\alpha_{T}$
as a function of the K/$\pi$ ratio for the MINOS data.
The errors in the experimental and theoretical values of
$\alpha_T$ are taken to be $\pm$~0.012 and $\pm$~0.013 respectively,
obtained by combining the statistical errors in quadrature
with the systematic errors in Tables~\ref{tab:errors}
and~\ref{tab:therrors}, but omitting
the error in the K/$\pi$ ratio in each case.  The error on the
theoretical value of $\alpha_T$ grows with increasing K/$\pi$ ratio
because $\epsilon_K$ has a larger uncertainty than $\epsilon_{\pi}$, so
a larger contribution from kaons introduces more uncertainty.  The intersection
of the two curves occurs at K/$\pi$~=~$0.12^{+0.07}_{-0.05}$.
The uncertainty is estimated
by assuming Gaussian errors for the the
theoretical and experimental values of $\alpha_T$ and performing a
$\chi^2$ minimization to determine the 
$\Delta \chi^2 = 1$
contour that encompasses the best fit point. 

\begin{figure}[!h]
\begin{center}
\includegraphics[width=.5\textwidth]{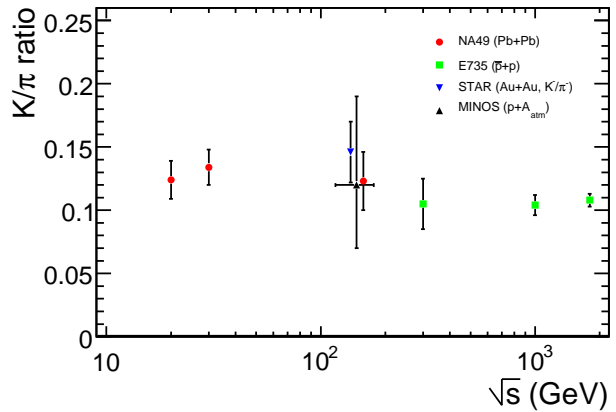}
\end{center}
\caption[ Summary of  K/$\pi$ Measurements]{\label{fig:kpicomp} A
compilation of selected measurements of $K/\pi$  for various center of
mass energies.  The STAR value was from Au+Au collisions at
RHIC~\cite{Adler:2002wn}, the NA49 measurement was from Pb+Pb collisions
at SPS~\cite{Afanasiev:2002mx,Alt:2005zq}, and the E735 measurement was
from p+$\bar{\mathrm{p}}$ collisions at the Tevatron~\cite{Alexopoulos:1993wt}.
 } 
\end{figure}

Previous measurements of the K/$\pi$ ratio do not directly compare to this
indirect measurement.  Nevertheless, the central value of MINOS's
measurement is consistent with the collider-based direct measurements,
although the indirect error bars are larger than those on
the direct measurements.  
A comparison of this measurement to other measurements is shown in
Fig.~\ref{fig:kpicomp}.  
Only the MINOS result is
for a reaction where the interacting particles do not have equivalent
energy in the laboratory frame.  
Nevertheless, they are all presented on the same
axes for a broad overview.
The central value of MINOS' indirect cosmic
ray-based $K/\pi$ measurement is consistent with the collider-based direct
measurements, and the associated error bars span the dispersion in
those direct measurements.

\section{Conclusions}
A five year sample of 67.32 million cosmic ray induced muons has been
collected by the MINOS far detector and daily rate fluctuations have
been compared to daily fluctuations in atmospheric temperature.
These distributions were shown to be highly correlated, with a
correlation coefficient of 0.90.  The constant of proportionality
relating the two distributions, $\alpha_T$, was found to be
0.873~$\pm$~0.009~(stat.)~$\pm$~0.010~(syst.).  This value is in good agreement with the
theoretical expectation
of $\left<\alpha_T\right>~=~0.864~\pm~0.024$.
 A measurement of the temperature dependence of the rate of $\mu^+$
separate from $\mu^-$ was performed for the first time.   There is no
statistically significant difference between $\alpha_T(\mu^+)$ and
$\alpha_T(\mu^-)$.  

The experimental value of $\alpha_T$ for the combined muon sample  has the
lowest uncertainty of any such measurement.  While other experiments
have estimated the effect of atmospheric temperature on kaon induced
muons~\cite{Barrett:1952,Ambrosio:1997tc},
this is the first result to quantify the kaon-inclusive effective temperature
coefficient.
The new kaon-inclusive model fits the MINOS far detector data better than the pion only
model~\cite{Ambrosio:1997tc} and
suggests a measurement of the atmospheric  K/$\pi$ ratio.
Applying the differing temperature variations of kaon and pion decay to
the seasonal variations analysis allowed the  first measurement of the
atmospheric K/$\pi$ ratio for $E_{p}\gtrsim$\unit[7]{TeV}.  It was found to be
K/$\pi$~=~$0.12^{+0.07}_{-0.05}$.

\section{Acknowledgments}
We thank the Fermilab staff and the technical staffs of the participating institutions for their vital contributions. This work was supported by the U.S. Department of Energy,  the U.K. Science and Technologies
Facilities Council, the U.S. National Science Foundation, the State and University of Minnesota, the Office of
Special Accounts for Research Grants of the University of Athens,
Greece, FAPESP (Fundacao de Amparo a Pesquisa do Estado de Sao Paulo) and
CNPq (Conselho Nacional de Desenvolvimento Cientifico e Tecnologico) in
Brazil. We gratefully acknowledge the Minnesota Department of Natural
Resources for their assistance and for allowing us access to the
facilities of the Soudan Underground Mine State Park and the crew
of the Soudan Underground Physics Laboratory for their tireless work in
building and operating the MINOS far detector.  We also acknowledge the BADC
and the ECMWF for providing the environmental data for this project.


\end{document}